# A new XML conversion process for mensural music encoding : CMME_to_MEI (via Verovio)

David Fiala

University of Tours (F)

david.fiala@univ-tours.fr

Laurent Pugin

RISM Digital Center, Bern (CH)

laurent.pugin@rism.digital

Marnix van Berchum

Huygens Institute for History and Culture of the Netherlands (NL)

marnix.van.berchum@huygens.knaw.nl

Martha Thomae

NOVA University of Lisbon (PT)

marthathomae@fcsh.unl.pt

Kévin Roger

University of Lorraine (F)

kevin.roger@univ-lorraine.fr

# Introduction

In 2020, the Ricercar Lab—the musicological research team at the Center for Advanced Studies in the Renaissance, University of Tours—took new steps managing its digital resources accumulated since the 1990s. Its first objective was to create a centralised online database, the [Ricercar Data Lab](#) (RDL), ready to launch publicly in June 2025 ([Vendrix 2024](#)).

While designing RDL, the team collaborated with two French Digital Humanities infrastructures, [Biblissima](#) and [Huma-Num](#), shaping the concept of digital musicology in the wider context of the latest interoperable tools and practices for studying heritage and humanities. These activities produced, for example, reflections and developments on the alignment of music books' digitisations with musicological metadata ([Fiala & Roger 2023](#)) and led to the search for specific solutions for the production and open-access publication of encoded early music.

Past projects of the Ricercar Lab ([CRIM](#), [Gesualdo Online](#)) have repeatedly shown that digital music editions were very costly to produce ([Vendrix 2023](#)) but also the crucial foundation for any ambitious realisation in digital musicology. It is in this context that an essential corpus was identified: the editions published by Clemens Goldberg on his foundation's website from the early 2010s onwards ([Goldberg Online](#)).

# 1. The Golberg Stiftung online Editions

A specialist of $15^{th}$-c. French music, Goldberg encoded the full content of 34 major music sources of this time, totaling 3356 files. Another 260 complementary files allowed him to offer series of complete collections, such as the complete works of Okeghem, the complete chansons of all contemporaneous major composers (Du Fay, Binchois, Caron, Busnoys, Ghizeghem, Compère, Agricola…), and the complete reworkings of most famous chansons (*De tous biens plaine*, *D'un aultre amer*…). Such a corpus is of major interest in itself, but all the more so for a research institution located a few blocks from Okeghem's house, with specific interests in the French chanson and the musical life in the Loire valley.

Yet another feature of Goldberg's editions was of particular interest to the Ricercar Lab. Although the Goldberg Foundation website only offers the option to download PDF files of its series of collections, these editions have been encoded in an XML format named CMME, which the Ricercar Lab had contributed to develop 20 years ago. The lab itself used it for encoding 5000 musical incipits for its catalogue of $16t^{h}$ c. chansons[1].

All these elements led the Ricercar Lab, with Biblissima's support, to organise and fund 1. the purchase of the original CMME files owned by the Goldberg foundation and 2. the development of a set of conversion tools for CMME files to meet more up-to-date standards of music encoding, namely MEI. This article focuses on this last development which opens new perspectives for the encoding of early music.

# 2. The CMME project and its components

CMME (Computerized Mensural Music Editing) was conceived by Theodor Dumitrescu as an undergraduate project at Princeton University in 1999. After one year at the CESR, Tours (2005-2006)—where the basics for the XML format, the transcribing/editing/viewing software and web environment were laid—the project moved to Utrecht University with a three-year funding by NWO (Dumitrescu & Berchum 2009). In the years at Utrecht, the project team worked on a first collaborative editorial project of the Occo Codex. In 2011, CMME received a small grant (SURF) to work on the new editorial project "The Other Josquin" and prepare the CMME environment for the emerging semantic web. After 2011, the project received no further funding and entered a dormant state.

---

[1] See [first item of this catalogue (older interface)](#).

At the end of its funded period, CMME made all of the developed software, editions and metadata available as open access. The 'stack' consist of a custom CMME-XML format, tailored to mensural notation; software for transcribing/editing mensural music (the 'Editor') and viewing the scores (the 'Viewer'); and an online environment for publishing the Editorial Projects, individual scores and the related metadata.[2] Due to changing web policies, the online version of the viewer software, integrated into the editorial projects, became unsupported by browsers. This made the usage of the online material far from ideal and demonstrated the urgency of moving to, or at the least enabling the conversion to, other music encoding standards.

## 3. Bringing CMME back into the MEI environment

With Biblissima's support, a workshop was organised in September 2024 at the Campus Condorcet in Paris, gathering a small team of experts with a wide range of knowledge: CMME, mensural notation, MEI mensural module, general MEI ecosystem, programming skills. Their goal was to put in place a set of tools for converting CMME files to MEI.

As the project progressed, the decision was made to integrate the converter directly into Verovio. Indeed, this open-source rendering library for MEI developed by the RISM Digital Center in Bern (Pugin, Roland & Zitellini 2014) supports both mensural notation and CMN rendering and acts as a converter to MEI for other important music notation formats (e.g., Humdrum and MusicXML). Furthermore, Verovio is available in different bindings, including JavaScript and Python, which means offering out-of-the-box a whole range of environments in which the converter can be used or embedded. An additional argument was the use of LibMEI, which directly ties the tool with the MEI Schema, facilitating the implementation and maintenance of the tool. The conversion from CMME to MEI mensural was implemented during the workshop, while the conversion to MEI CMN was implemented afterwards by the RISM Digital Center.

Implementing the converter confirmed that MEI for mensural notation is already quite mature and can represent almost everything that CMME does. However, the mapping from CMME to MEI is not always straightforward and the conversion cannot always rely on a one-to-one correspondence. Not surprisingly, durations are the most difficult to handle. A new option had to be introduced to Verovio for replicating the CMME equivalence duration on the minima. For proportions, a well-known complexity of mensural music, CMME makes the distinction between actual proportions, and proportions representing tempo changes. Since there is no such distinction in MEI, typing the proportion with a generic attribute (@type) was necessary. Also, not all cases would be precisely disentangled semantically even though the converter always produces the same voice alignment.

---

[2] The source code, including the xsd-file of the CMME-XML format, is available at https://github.com/tdumitrescu/cmme-editor; the website is www.cmme.org; all CMME-XML scores produced by the project are available at https://github.com/tdumitrescu/cmme-music.

Using a generic typing attribute was used in quite a few places in the conversion to MEI (notes marked as *signa congruentiae*, gaps marked as ellipsis, and variants marked as lacuna). For all these cases, the use of the generic typing attribute allows for the information to be preserved, although not in a fully satisfactory manner. These will eventually be considered on a case-by-case basis by the MEI mensural-ig to improve the semantic structure for the conversion project. Along the same lines, a few missing features in MEI were identified (e.g., support for accidentals within ligatures) and will be potentially proposed as additions for the next version of MEI by the mensural-ig.

The conversion to CMN in Verovio is triggered with a new option, available not only for CMME files but also for mensural MEI files. In the conversion to CMN, the editorial markup is dropped. However, using the editorial markup selection option in Verovio, it is possible for a particular reading to be selected for the conversion to CMN. Mensural-specific features such as ligatures and coloration are also dropped during the conversion process, but preserved as brackets and bracket angles, following standard practices. Clefs are normalised to G2, G2 with *ottava bassa*, and F4. All note durations are preserved with a mapping of a *minima* to a half-note and with the longest duration being the *breve*—tied notes are then used to represent longer durations. In the barring process and when generating measures, meter signatures are selected to best accommodate the mensur signs. More precisely, mensur signs are converted into CMN meter signatures 3/2, 4/2 and 6/2. The meter signature is arbitrarily determined by the top voice when voices have conflicting meter signatures. Also, a meter signature change is introduced when all voices have a mensur sign at the position and the resulting meter signature is actually different from the previous one. The mensur signs are not preserved as such. However, a MEI dir (directive) is inserted with a textual transcription representing the mensur sign. Finally, proportions are converted into tuplets. A tuplet that matches the proportion can be used when notes do not need to be cut across several bars. Otherwise, the tuplet proportion needs to be adjusted according to the proportion of the note part that goes in one bar and the other. The result is more or less legible but mathematically correct.

Overall, the result of the conversion to CMN is meant to be as simple as possible. The conversion supports the MEI-Basic option, which makes the data widely reusable. The files produced by the converter with the MEI-Basic option enabled can thus be loaded into MuseScore, opening up a wide range of use-case scenarios for the CMME data.[3]

# Conclusion

With the availability of a direct import of CMME-XML into Verovio, the corpus of existing CMME files gets a new life. Furthermore, since the stand-alone CMME software still works fine and no alternative is available for native MEI, one can imagine a pipeline for encoding and editing mensural notated music. A first transcription, including variant readings, can be made with CMME, after which the MEI software components take care of conversion, further edition (e.g. in mei-friend) and publication (Verovio).

---

[3] The only exception are complex proportions that can, in some cases, produce tuplets that are appropriately represented in MEI but not supported by MuseScore, but that is only for a very small portion of the files the converter has been tested with.

## Acknowledgements

The authors acknowledge the support of Biblissima+, an infrastructure funded by the French government and managed by the ANR for the Programme d'investissements d'avenir/France 2030, ref. ANR-21-ESRE-0005.